# Spectro-interferometry of the Be star $\delta$ Sco: Near-Infrared Continuum and Gas Emission Region Sizes in 2007

R. Millan-Gabet

California Institute of Technology, NASA Exoplanet Science Institute, Pasadena, CA 91125, USA

R.Millan-Gabet@caltech.edu

J. D. Monnier

Department of Astronomy, University of Michigan, MI 48109, USA

Y. Touhami and D. Gies

Center for High Angular Resolution Astronomy and Department of Physics and Astronomy, Georgia State University, Atlanta, GA 30302, USA

E. Hesselbach

University of Toledo, Department of Physics & Astronomy, Toledo, OH 43606, USA

E. Pedretti and N. Thureau

School of Physics and Astronomy University of St Andrews, St Andrews, Fife, KY16 9SS, Scotland, United Kingdom

M. Zhao

Jet Propulsion Laboratory, 4800 Oak Grove Drive, M.S. 169-327, Pasadena, California 91101, USA

and

T. ten Brummelaar and the CHARA Group

Center for High Angular Resolution Astronomy, Georgia State University, PO Box 3969, Atlanta, Georgia 30302-3969, USA

## ABSTRACT

We present near–infrared $H$ and $K$–band spectro–interferometric observations of the gaseous disk around the primary Be star in the $\delta$ Sco binary system, obtained in 2007 (between periastron passages in 2000 and 2011). Observations using the CHARA/MIRC instrument at $H$–band resolve an elongated disk with a Gaussian FWHM $1.18 \times 0.91$ mas. Using the Keck Interferometer, the source of the $K$–band continuum emission is only marginally spatially resolved, and consequently we estimate a relatively uncertain $K$–band continuum disk FWHM of $0.7 \pm 0.3$ mas. Line emission on the other hand, He I $\lambda 2.0583$ $\mu$m and Br$\gamma$ $\lambda 2.1657$ $\mu$m, is clearly detected, with $\sim 10\%$ lower visibilities than those of the continuum. When taking into account the continuum/line flux ratio this translates into much larger sizes for the line emission regions: $2.2 \pm 0.4$ mas and $1.9 \pm 0.3$ mas for He I and Br$\gamma$ respectively. Our KI data also reveal a relatively flat spectral differential phase response, ruling out significant off–center emission. We expect these new measurements will help constrain dynamical models being actively developed in order to explain the disk formation process in the $\delta$ Sco system and Be stars in general.

*Subject headings:* techniques: high angular resolution — stars: emission-line, Be — stars: individual ($\delta$ Sco)

## 1. Introduction

Be stars are massive, rapidly rotating stars, on or near the main–sequence, which, at least at some time during their evolution, exhibit peculiar observational characteristics: infrared excesses compared to photospheric levels, emission lines, partial polarization of radiated light, and variability of all those properties on many time–scales, from weeks to decades. It is generally understood that the observational characteristics that define Be stars arise from the development of a thin disk of gas in the equatorial plane. The precise physical processes that govern its formation are, however, not well understood. Processes that are believed to dictate or participate in the formation of a gaseous equatorial disk include: equatorial mass–loss due to near–critical rotation, radiatively drive winds, and photospheric pulsations (see e.g. review by Porter & Rivinus (2003). The role, if any, that binarity plays in the Be phenomenon is also unclear, although many Be stars may have been spun up in the past by mass exchange (Peters et al. 2008).

The star $\delta$ Sco (HD 143275, HIP 78401, HR 5953; V $\sim$ 2.3, H $\sim$ 2.4, K $\sim$ 2.4) was until recently classified as an unremarkable early–B type star. It is also a highly eccentric binary system with a 10 year period and $\sim$ 0.1 arcsec angular separation (Bedding 1993; Mason et al. 2009; Tango et al. 2009). First signs of its Be nature were observed by Coté & van Kerkwijk (1993), and a dramatic photometric and activity outburst was discovered shortly following its 2000 July periastron passage (Otero et al. 2001). The possible link between the close periastron passages and the onset of activity was suggested early on (Bedding 1993) and was certainly apparent during the 2000 periastron passage (Halonen et al. 2008). In terms of average properties, it is found that the post–outburst properties of $\delta$ Sco are similar to those of typical Be stars, and are satisfactorily explained with models of an equatorial gaseous disk (Banerjee et al. 2001; Miroshnichenko et al. 2001; Carciofi et al. 2006). Many intriguing and unexplained features remain however, in particular pertaining to the variability, over several distinct time–scales, of the photometric and spectroscopic activity indicators.

Here we present new observation of the δ Sco primary, obtained in 2007, seven years after the 2000 periastron passage and three years before the following one in 2011. Our spectro–interferometric observations spatially resolve distinct sources of emission: the $H$ and $K$–band continuum (star + gaseous disk), and regions of He I and Brγ emission.

## 2. Observations and Observables

### 2.1. CHARA/MIRC

Observations of δ Sco were made on UT 2007 May 10–12 (JD 2,454,231–233) at the CHARA interferometer (ten Brummelaar et al. 2008), using the Michigan InfraRed Combiner instrument (MIRC, Monnier et al. (2008)). MIRC operates in $H$–band ($\lambda_0 = 1.65\mu$m, $\Delta\lambda = 0.28\mu$m). For these observations we used a mode which disperses the image plane fringes over 8 spectral pixels, in order to avoid bandwidth–smearing coherence losses We used four CHARA telescopes resulting in six simultaneous baselines, with the following average projected lengths and position angles (measured East of North): E1W1 (301.7 m, 203°.2), S1E1 (283.3 m, 44°.7), E1W2 (220.9 m, 206°.8), S1W1 (124.0 m, 136°.9), S1W2 (103.2 m, 94°.3), and W1W1 (86.1 m, 12°.6).

Following standard practice, observations of δ Sco were interleaved with observations of calibrator stars used to monitor the instrument's transfer function. The calibrators used were: HD190327 (V=5.5, H=3.3, K0III, $\theta = 1.097\pm0.014$ mas), HD149757[1] (V=2.6, H=2.7,

---

[1]Note that HD149757 = zeta Oph is sometimes a Be star with H-Balmer emission. However, according to the Database of Be Star Spectra (http://basebe.obspm.fr/basebe/) ob-

O9V, $\theta = 0.51 \pm 0.05$ mas), and HD164259 (V=4.6, H=3.7, F2IV, $\theta = 0.69 \pm 0.03$ mas); where their uniform disk angular diameters ($\theta$) and their errors were estimated using the surface brightness relations of Barnes et al. (1978).

Reduction and calibration of the MIRC data were performed using its standard pipeline, as described in Monnier et al. (2007). While an analysis of multi–epoch closure phases is under way (including the effects of the companion orbital motion and investigation of possible off–center disk emission), for the purposes of the work presented here we only concern ourselves with the visibility amplitudes, in order to measure the $H$-band size of the disk surrounding the primary star.

Using the most recent orbital solutions (Mason et al. 2009; Tango et al. 2009), we compute that the angular separation of the $\delta$ Sco companion to be 182 mas at this epoch. This is inside the MIRC field of view (FOV $\sim$ 0.5 arcsec, approximately matched to the CHARA telescopes diffraction limit). However, for an interferometer, the FOV is also limited by the bandwidth coherence envelope (see e.g. Lawson (2000)), which for MIRC in this spectral resolution mode ranges from 45 mas for the longest projected baseline to 159 mas for the shortest projected baseline.

Figure 1 shows the uv coverage for the CHARA/MIRC observations and the calibrated visibility data. The system appears clearly resolved on all baselines (visibility < 1.0). We note that, in this 1–dimensional plot, strong object assymetries appear as variations in the visibilities at a given value of the projected baseline, and this, and not data noise, is the source of much of the variations seen at short baselines, due to the effect of the companion (as indicated above, the effect of the companion decreases and then completely

---

servations of zeta Oph made on 2007 March 26 and 2007 July 14 show H-alpha as a pure absorption line. Consequently, it was probably safe to use as a calibrator in 2007 May.

disappears at longer baselines as the coherence envelope decreases). The errors shown include a systematic term arising from calibration of seeing. Note that the relatively southern declination results in limited uv coverage along most position angles, which will limit our ability to constrain the position angle of the disk surrounding the primary star.

## 2.2. Keck Interferometer

Observations of $\delta$ Sco were made on UT 2007 July 03 (JD 2,454,285) at the Keck Interferometer (KI, Colavita et al. (2004)). We used a low spectral dispersion mode, which provides 42 spectral channels across the near–infrared $K$–band ($\lambda_0 = 2.18\mu$m, $\Delta\lambda = 0.4\mu$m, $R = 200$, see Eisner et al. (2007)). The width of each spectral channel is therefore $\sim 93$Å. The average projected length and position angle of the KI baseline was 77.5 m and 44°.2 respectively. The calibrator stars used were: HD124850 (V=4.1, K=2.8, F7IV, $\theta = 1.15 \pm 0.07$ mas) and HD135742 (V=2.6, K=2.9, B8V, $\theta = 0.69 \pm 0.05$ mas).

Reduction and calibration of the interferometer data (visibility amplitudes, differential phases and spectral fluxes) were done using standard packages provided by the NExScI[2]. The flux spectra (also obtained from the interferometer data) were normalized to the flux in the longest wavelength channel (2.39$\mu$m), and corrected by dividing by the average of the normalized calibrator spectra. The spectral differential phase data are referenced to the phase of the broadband $K$–band channel (used to "fringe track", i.e., to stabilize the interference fringes against the optical path differences caused by the atmosphere and instrument). Two independent observations of $\delta$ Sco yield $V^2$ that agree to better than 1%. Nevertheless, per project documentation describing the performance of the KI instrument, we added an external error of 0.03 to the formal statistical errors in the absolute calibration

---

[2]i.e., *Kvis* and *nbCalib*; http://nexsci.caltech.edu/software/KISupport/

of $V^2$.

The angular separation of the δ Sco companion was 180 mas at the epoch of the KI observations, much larger than the 50 mas field FOV of the KI near–infrared instrument (set by the acceptance angle of the single–mode fiber which feeds the fringe tracker camera, and also matched to the KI telescopes diffraction limit) . The companion therefore adds no coherent or incoherent flux, and we may ignore its existence for the purposes of interpreting the KI data.

The calibrated visibility amplitude, differential phase and flux data are shown in Figure 2. The following are the main features one can see in these data: (1) Spectral features are clearly seen in the visibility and flux data, attributed to He I ($2.0583\mu$m) and Brγ ($2.1657\mu$m) gaseous line emission. We note that He I and Brγ emission were also observed in $K$-band spectra made shortly after after the last periastron brightening (Banerjee et al. 2001). (2) The visibilities at the wavelengths of these lines are *lower* than that at wavelengths of continuum–only emission (by $\sim 10\%$), indicating that the line emission region is *larger* than the continuum emission region (actually, by a large factor, as will be seen in the next section). (3) The $K$–band continuum is only slightly resolved ($\overline{V^2} = 0.95 \pm 0.03$). (4) The differential phase spectrum is featureless, ruling out significant off–center emission. We note that the KI spectral resolution in this mode does not spectrally resolve the emission lines (for example, for a classical Be star high rotation rate of 400 km s$^{-1}$, we expect a FWHM(Brγ) $\sim$ 58Å, or ×1.6 narrower than the KI spectral resolution). The fact that the lines in the flux and visibility spectra of Figure 2 appear to span more than one pixel is simply because the wavelengths of line emission are not centered on the spectral pixels.

The calibrated CHARA and KI data are available in the standard format for optical interferometer data (OI-FITS, Pauls et al. (2005)) at: http://olbin.jpl.nasa.gov/data/.

## 3. Sizes of the Continuum and Line Emission Regions

### 3.1. $H$–band Continuum (CHARA/MIRC)

We fit the CHARA/MIRC observations with a model consisting of a central primary star, an elliptical Gaussian brightness (representing the gaseous circumstellar disk), and a companion star. For equations describing the interferometer's response to these standard morphologies, see e.g. Berger & Segransan (2007). Here and in the following sections we choose a Gaussian brightness because it has been shown in simulations (e.g. Stee et al. (1995)) and experimentally (Tycner et al. 2006) to be a good approximation for the disk brightness (and much better than e.g. a uniform disk or ring).

For the primary photosphere, we adopt a uniform disk angular diameter of $0.51 \pm 0.06$ mas. This value represents an average of estimates obtained by two different methods: the pre–active primary stellar parameters of Carciofi et al. (2006), and the surface brightness–color relations of Barnes et al. (1978), and the 12% uncertainty represents the scatter among the various estimates. For the companion, we estimate a uniform disk angular diameter of $0.27 \pm 0.03$ mas, by following Carciofi et al. (2006) in adopting an early B spectral type and a secondary to primary flux ratio at all wavelengths of $F_s/F_p = 0.16 - 0.25$ (based on the measured visual magnitude difference; Bedding (1993); Tango et al. (2009)), and again using the Barnes et al. (1978) relations.

Thus, the free parameters in our fit are the primary to total $H$–band flux ratio $(F_p/F_T)_H$, the elliptical Gaussian brightness FWHM along the major and minor axis (FWHM$_M$ and FWHM$_m$), its orientation on the sky (position angle, PA$_M$), and the companion coordinate offsets. The baseline–dependent bandwidth smearing effect on the signature form the companion is fully accounted for in the modelling. The best–fit solution places the companion at a separation of $\sim 190$ mas, consistent with expectations from the orbital solution. However, because the long baselines contain no information about the companion, the data cannot constrain its position angle (as indicated by the fact that fitting the different epochs separately yields very different results). We note however that this does not affect the robustness of the fit to the disk parameters.

The best fit solution (Table 1) has $(F_p/F_T)_H = 0.57 \pm 0.05$, $\text{FWHM}_M = 1.18 \pm 0.16$ mas and $\text{FWHM}_m = 0.91 \pm 0.12$ mas. We note that in the thin disk approximation the ratio of minor to major axis sizes yield the inclination, and the MIRC fit implies $i = 39 \pm 17°$, consistent with $i = 38°$ from Carciofi et al. (2006). As can be seen in Figure 1, the $(u, v)$ coverage is limited along most directions (due to the Southern declination), resulting in a rather uncertain disk position angle $\text{PA}_M = 25 \pm 29°$.

### 3.2. $K$–band Continuum (KI)

In order to compare the angular size for the region of $K$–band continuum emission with that of the primary star alone, we define the spectral pixels outside the lines and fit the observed visibilities with a model consisting of a single uniform disk brightness. We obtain a mean continuum uniform disk diameter of $0.8 \pm 0.3$ mas. This represents only a marginal $\sim 1\sigma$ detection of a $K$–band "continuum size" in excess of what is expected from the primary photosphere alone ($\theta = 0.51 \pm 0.06$ mas, see also Figure 2). However, the detection is strengthened by the fact that the uniform disk diameters fit to the individual spectral pixels increase monotonically across the $K$–band, indicating that we resolve an underlying brightness of wavelength dependent radius. If the underlying brightness were simply that of the primary photosphere, and given that for this type of star and wavelength limb–darkening would be undetectable at this spatial resolution, the uniform disk diameters would be the same for all continuum spectral channels.

This "continuum size" includes both the central primary star and $K$–band continuum emission from the gaseous disk. In order to obtain a characteristic size for the $K$–band *disk* emission, we fit the continuum visibilities to a model consisting the central primary star plus a Gaussian brightness. The KI measurement was made at essentially a single spatial frequency; therefore we cannot constrain the disk geometry (inclination and position angle) and we use a face–on model (circularly symmetric). For the same reason, in order to fit the Gaussian FWHM we require an independent estimate of the primary to total $K$–band flux ratio $(F_p/F_T)_K$ via, for example, spectral energy distribution (SED) decomposition. Unfortunately, there exists no near–contemporaneous SED in the published or un–published literature (to our knowledge). We thus form an estimate using the $K$–band flux seen by the interferometer itself, calibrated by that measured during the calibrator observations. We obtain the total $K$–band magnitude $K_{\text{total}} = 2.38 \pm 0.19$. In order to estimate the primary star $K$–band flux, we use the pre–outburst photometry of The et al. (1986) which give a total K magnitude (primary plus secondary) of 2.70. Thus the magnitude of the primary alone is $K_{\text{primary}} = 2.90 \pm 0.04$ (considering the error in the secondary/primary flux ratio, as described in the previous section). Thus, we obtain $(F_p/F_T)_K = 0.62 \pm 0.12$. By fitting each continuum spectral pixel, we obtain a FWHM that monotonically increases from $0.4 \pm 0.4$ mas to $0.9 \pm 0.3$ mas across the $K$–band. The mean $K$–band FWHM is $0.7 \pm 0.3$ mas (Table 1). The uncertainties in the input parameters (primary to total $K$–band flux ratio and angular size of the primary star) introduce an additional 10% error in the continuum disk size.

### 3.3. Size of the Line Emission Regions

At wavelengths of line emission, there is also continuum emission. Assuming no off-center emission (consistent with our DP data), the measured visibilities as a function of wavelength may be decomposed as:

$$V_{\lambda i} = \frac{(F_{\lambda i,c} \cdot V_{\lambda i,c}(\theta_{\lambda i,c}) + F_{\lambda i,l} \cdot V_{\lambda i,l}(\theta_{\lambda i,l}))}{(F_{\lambda i,c} + F_{\lambda i,l})} \quad (1)$$

where $V$ is the visibility modulus, $\theta$ represents a characteristic angular size, $F$ is the flux, and $\lambda i$ refer to the KI spectral bins. The wavelength subscripts "$c$" and "$l$" refer to the regions where the continuum or line emission originate. From this, it can be seen that if significant flux arises in compact continuum regions, as is the case here, even a modest decrease in the measured visibility (e.g. $\sim 10\%$ in Figure 2) can correspond to a large difference between the sizes of the continuum and line regions. For example, in the limiting case that the system consists of an unresolved central star surrounded by a very large, completely resolved, line–emitting region ($V_{\lambda l}(\theta_{\lambda l}) = 0.0$), the measured visibility in the line will not be zero, but given by the the fraction of flux coming from the continuum region: $V_\lambda = F_{\lambda c}/(F_{\lambda c} + F_{\lambda l})$. Within each spectral bin, we use Equation 1 monochromatically, at the center wavelength of each bin.

In the above equation, $\theta_{\lambda i,c}$ is the continuum size derived in the previous section; and before we can fit the line emission region sizes, we must estimate the contribution to the total flux from the continuum at each wavelength. This can be readily done using the flux spectrum measured by KI and interpolating the continuum flux under the lines. We emphasize that this is all that is required in order to obtain absolute sizes for the line emitting regions: determining the line sizes does not depend on the decomposition of the continuum emission into star+disk presented in the previous section, and therefore does not depend on our estimates of the primary star diameter or its contribution to the total flux.

As before, the line emitting gaseous disk is represented by a circularly symmetric Gaussian brightness distribution and we fit the FWHM. The results are shown in Figure 3, which shows the disk FWHM as a function of wavelength. As noted above, the line emission is much larger than the continuum size, ×3.2 and ×2.8 for He I and Brγ, respectively. Table 1 also summarizes these results.

These line region sizes may be considered upper limits, because we may be under–estimating the emission line fluxes if there is some line absorption due to the the primary star photosphere. We note that there appears to be no Brγ absorption in the spectrum shown in Banerjee et al. (2001) and indeed it is possible that the disk component completely obscures the photospheric component in $\delta$ Sco. Nevertheless, in order to quantify this effect, we use a Kurucz model (Kurucz 1979) for the primary star, for the stellar parameters of Carciofi et al. (2006). The photospheric Brγ absorption spans $\sim$ 2.4 KI spectral pixels, and amounts to $\sim$ 4.5% of the stellar continuum flux; or $\sim$ 2.25% of the total $K$–band continuum (recall that we estimated above that the primary star contributes 62% of the $K$–band continuum seen by KI). When this maximum correction is applied to the line/continuum flux ratio, we obtain a Brγ FWHM that is 4% smaller; a relatively small effect compared to the $\sim$ 15% errors in the sizes that result from the errors in the visibility data. The He I photospheric absorption is weaker, and the effect would be even smaller. The errors in the KI spectrum fluxes translate into $\sim$ 10% errors in the line/continuum flux fractions, which in turn translate into $\sim$ 30% errors in the fitted Gaussian FWHM.

## 4. Discussion

In Figure 4 we represent schematically the sizes measured for the different emission regions. Interestingly, within errors, we do not detect a significant difference between the $H$ and $K$–band continuum sizes. It remains to be seen whether this is generally the case for Be disks (preliminary results indicate that this is also the case for $\zeta$ Tauri, G. Schaefer, private communication).

In order to compare our new NIR measurements with predictions at other wavelengths, we make use of the relationship between H$\alpha$ luminosity and disk major axis empirically established by Tycner et al. (2005). We used near-contemporaneous spectra (UT 2007 July 12) obtained at the Ritter Observatory. We measure an EW($H\alpha$) = $-11.6$Å, which after correction for photospheric absorption (EW = 3.2Å, using the same Kurucz model as in Section 3.2), results in a net EW($H\alpha$) = $-14.8$Å. Following Tycner et al. (2005), this translates into an H$\alpha$ luminosity of $106.8 \times 10^{25}$ W and a disk semi-major FWHM of $72.4 \times 10^9$ m, or $14.9 R_\star$ (for $R_\star = 7 R_\odot$). This size scale is $\times 2.0$ and $\times 1.8$ bigger than the Br$\gamma$ and He I FWHM measured here along the KI position angle ($44°\!.4$ East of North). Although with large uncertainties, the CHARA/MIRC data constrains the semi–major axis position angle ($25 \pm 29°$) to be similar to that of the KI baseline, such that the size difference cannot be attributed solely to the projection of the KI baseline. Indeed, the optical depth in H$\alpha$ is likely to be much larger than in either of He I or Br$\gamma$, and therefore the disk will look much larger in H$\alpha$ (Gies et al. 2007). Furthermore, Carciofi et al. (2009) argue that Br$\gamma$ forms closer in than H$\alpha$ and is modulated in a different way by the one–armed spiral asymmetry in the disk.

Our measured line emission sizes in K–band are also $\sim \times 2$ smaller than the Carciofi et al. (2006) best SED–fitting solution (which implies $R_{\text{disk}} = 7 R_\star$). This radius however represents the outer boundary of the disk. The optical depth in Br$\gamma$ will be larger in the inner part of the disk where the density is higher, thus it is not surprising that we measure a smaller radius than the disk outer boundary. In any case, as pointed out by those authors, in those models disk size and density are degenerate. Our measurements provide direct constraints on the size, albeit for a later epoch.

We note that photometric light curves indicate that the state of the disk was less active (lower density) at the time of our observations compared to earlier years (see e.g. S. Otero's

– 15 –

web site `http://varsao.com.ar/delta_Sco.htm`); and our estimate of the primary + disk $K$–band magnitude ($2.38\pm0.19$) appears to connect with the 2005 $K$–band decline reported in Carciofi et al. (2006). The $H_\alpha$ emission was also on the decline (Pollmann 2009).

As pointed out above, several groups find that the $\delta$ Sco primary shows on average the observational signatures of a typical Be disk. But it has also been found that this system exhibits very unusual characteristics (e.g. Carciofi et al. 2006) which make it a very interesting case to study in detail. Observations near the periastron passages will offer us the opportunity to observe the onset of emission line activity, which can reveal fundamental information about the building of stellar envelopes, the causes of the Be phenomenon, and the role of binarity. Ultimately, dynamical models are needed and are being actively pursued by several groups. Our results presented here offer crucial observational constraints to those models.

The Keck Interferometer is funded by the National Aeronautics and Space Administration as part of its Navigator program. The data presented herein were obtained at the W.M. Keck Observatory, which is operated as a scientific partnership among the California Institute of Technology, the University of California and the National Aeronautics and Space Administration. The Observatory was made possible by the generous financial support of the W.M. Keck Foundation. The authors wish to recognize and acknowledge the very significant cultural role and reverence that the summit of Mauna Kea has always had within the indigenous Hawaiian community. We are most fortunate to have the opportunity to conduct observations from this mountain. The CHARA Array is funded by the National Science Foundation through NSF grants AST-0307562 and AST-0606958 and by the Georgia State University. J.D.M. acknowledges support from NSF grants AST-0352723 and AST-0707927.

*Facilities:* Keck Interferometer, CHARA interferometer, Ritter Observatory.

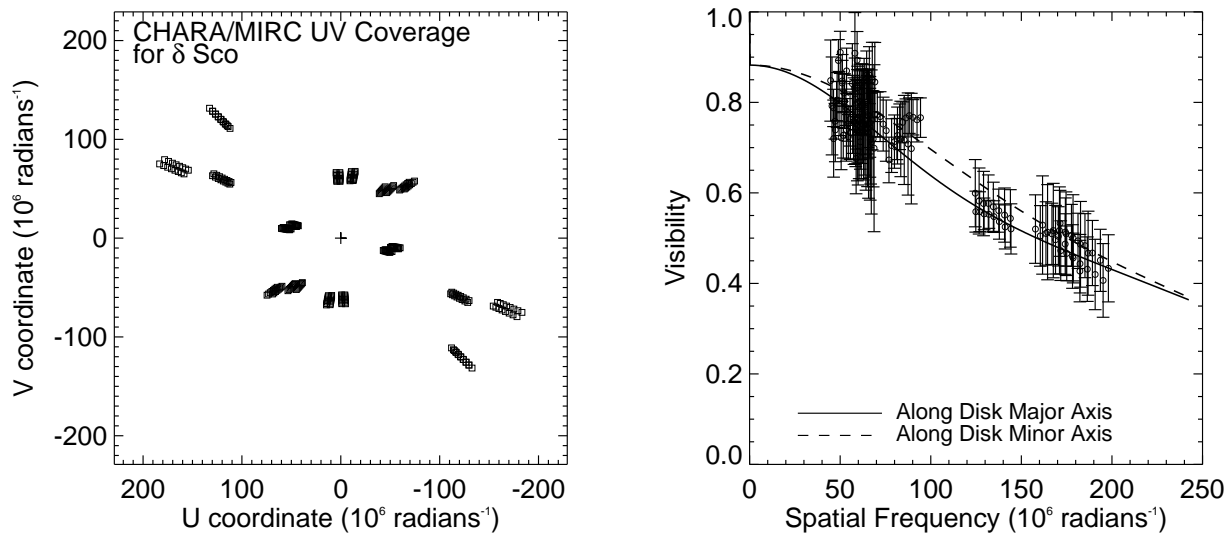

Fig. 1.— Observations made in the H–band using CHARA/MIRC. The left panel shows the uv coverage. The right panel shows the calibrated visibility data along with the best–fit visibility curve along the major and minor axes of the disk. Although the stellar companion is fully accounted for in the modelling, for clarity in this plot, we have removed it from the best–fit model shown. As described in the text, the variability in the visibilities seen at short baselines is due to the effect of the companion, the effect at longer baselines disappears because the coherence envelope becomes smaller than its angular separation.

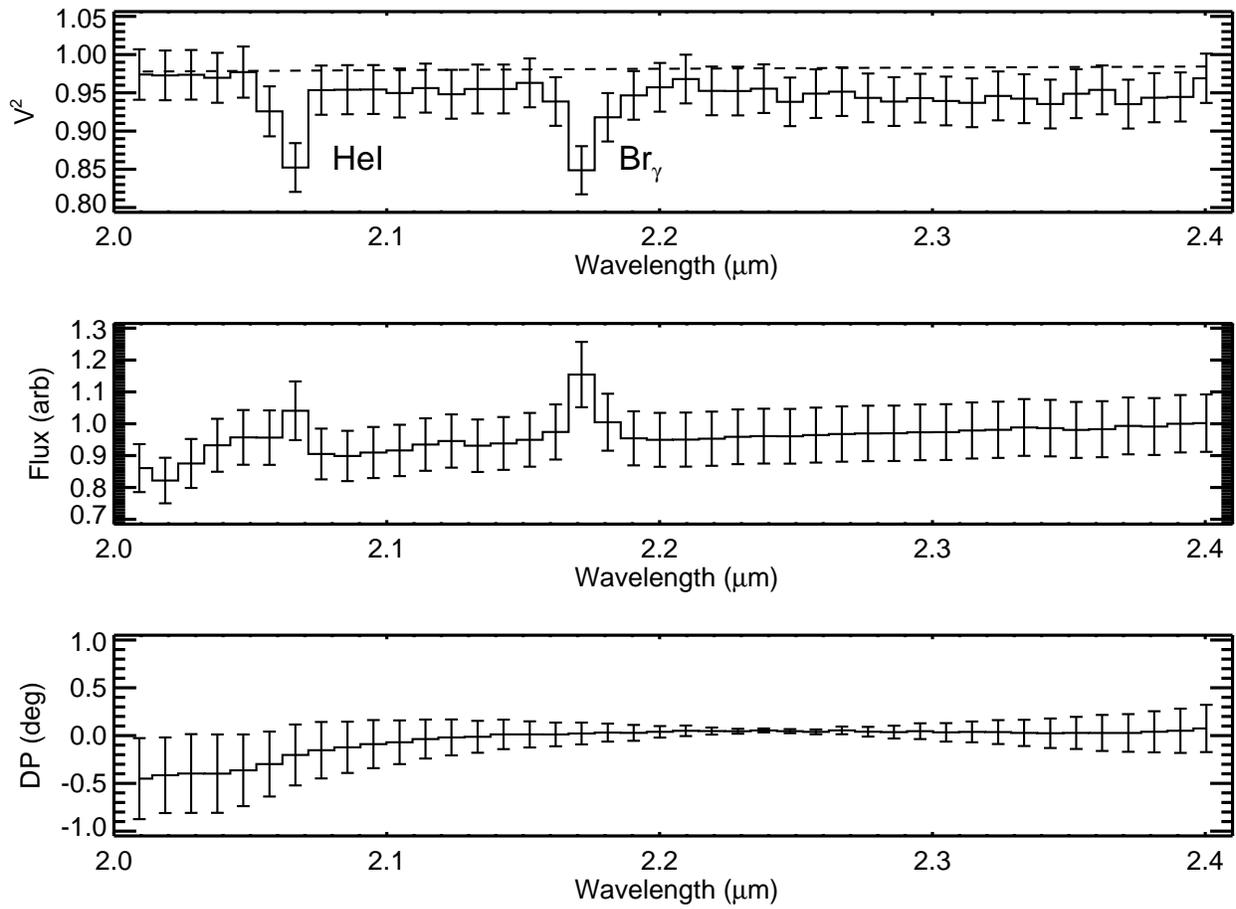

Fig. 2.— Keck Interferometer calibrated K–band data as a function of wavelength: $V^2$ (top), relative flux (middle), and differential phase (DP, bottom). The dashed line in the top panel represents the $V^2$ of the primary star photosphere (uniform disk diameter $0.51 \pm 0.06$ mas).

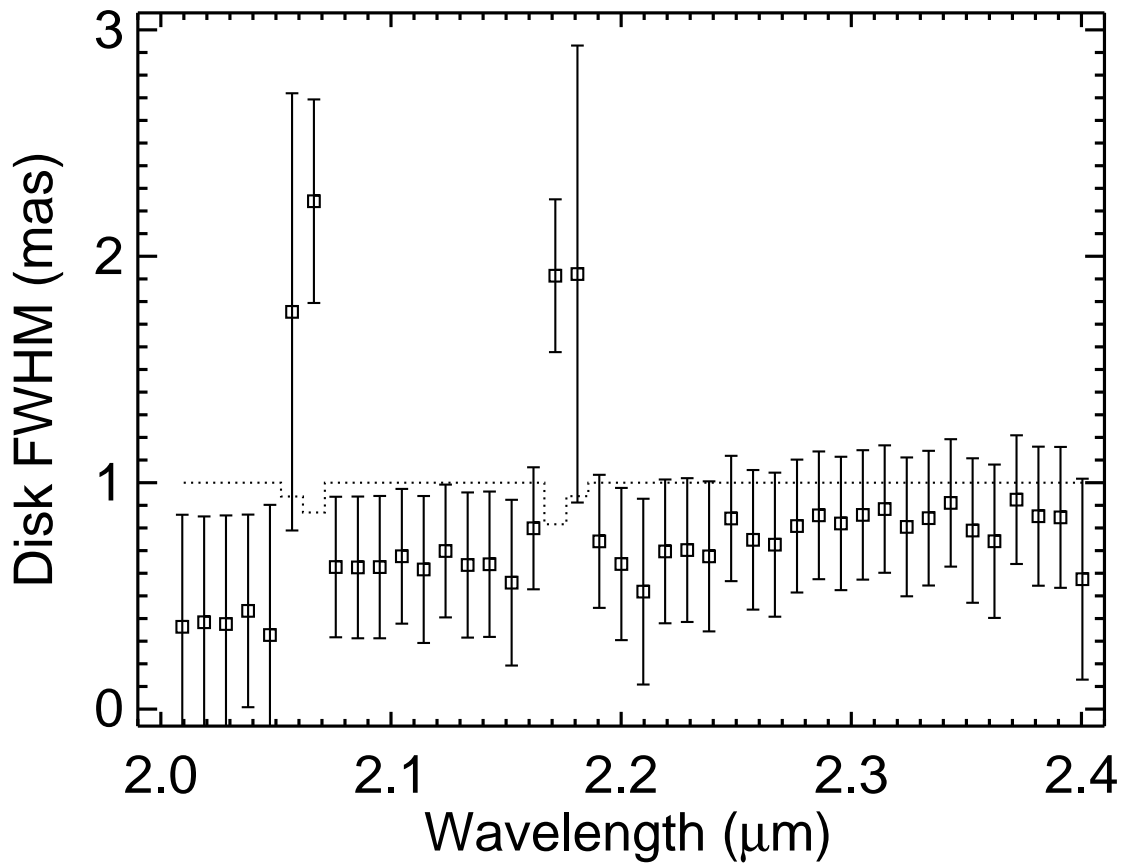

Fig. 3.— Gaussian FWHM fitted to the Keck Interferometer data for the continuum and line wavelengths. The dotted line represents the flux fraction (continuum/total), for the case that photospheric absorption is ignored. As discussed in the text, corrections for photospheric line absorption lead to line sizes that are at most 5% smaller.

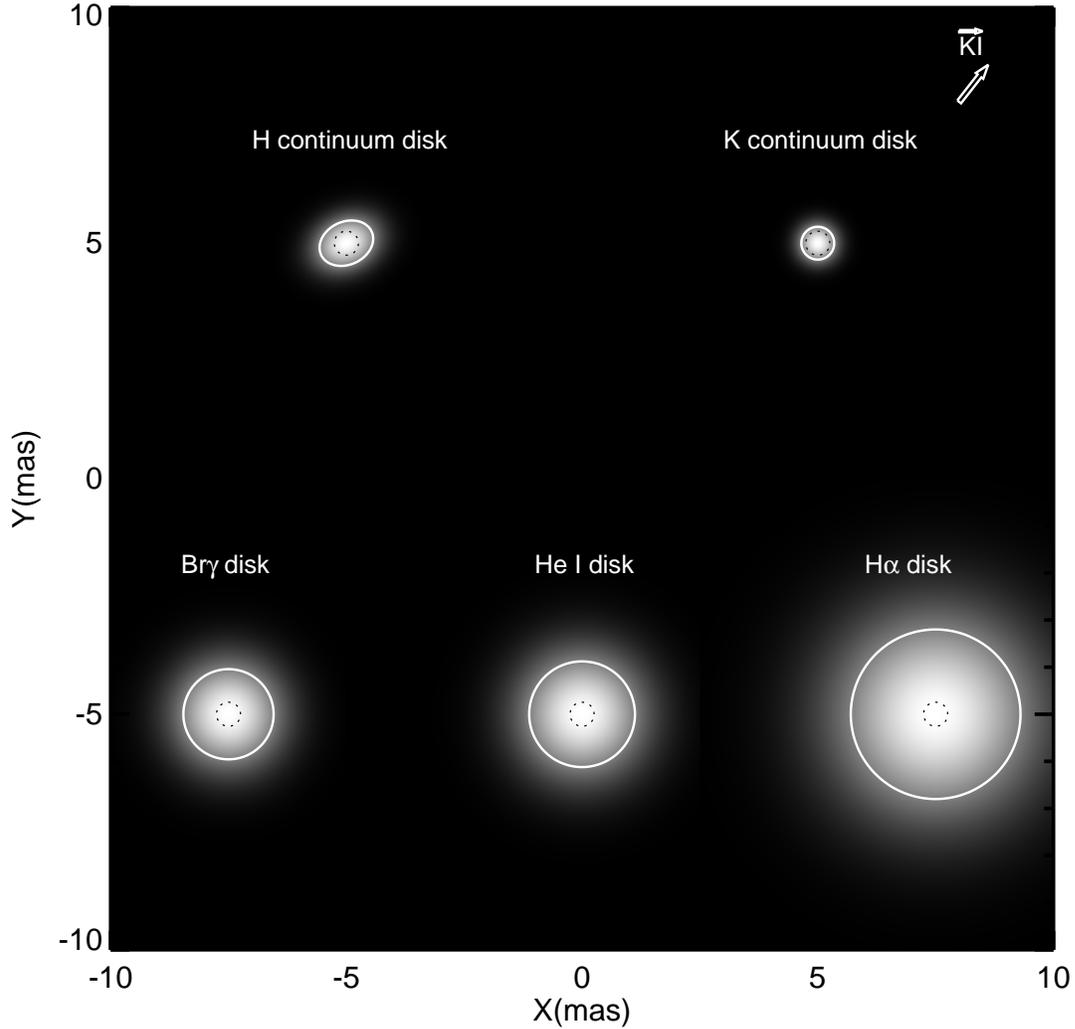

Fig. 4.— Representation of the disk FWHM for the various disk emission regions. The primary star is represented by the central dashed circles. For each emission region, we plot a Gaussian of the measured disk FWHM (bright circles). For $H_\alpha$, the size is not directly measured but inferred from the measured $H_\alpha$ spectrum, as described in the text. North is up and East is to the right.

Table 1: Results.

| | | | | | |
|---|---|---|---|---|---|
| | | H–band (CHARA/MIRC) | | | |
| | | Fixed parameters: | | | |
| | | $\theta_p = 0.51 \pm 0.06$ mas, $\theta_s = 0.27 \pm 0.03$ mas, $F_p/F_s = 0.16 - 0.25$. | | | |
| $(F_p/F_T)_H$ [a] | FWHM$_M$ | FWHM$_m$ | FWHM$_M$ | FWHM$_m$ | PA$_M$ |
| | (mas) | (mas) | ($D_\star$ [b]) | ($D_\star$) | (deg, East of North) |
| $0.57 \pm 0.05$ | $1.18 \pm 0.16$ | $0.91 \pm 0.12$ | $2.22 \pm 0.30$ | $1.71 \pm 0.22$ | $25 \pm 29$ |
| | | K–band (KI) | | | |
| | | Fixed parameters[c]: | | | |
| | | $\theta_p = 0.51 \pm 0.06$ mas, $(F_p/F_T)_K = 0.62 \pm 0.12$[a] | | | |
| Disk Region | | FWHM | | | FWHM |
| | | (mas) | | | ($D_\star$) |
| Continuum | | $0.7 \pm 0.3$ | | | $1.3 \pm 0.6$ |
| He I | | $2.2 \pm 0.4$ | | | $4.2 \pm 0.8$ |
| Br$\gamma$ | | $1.9 \pm 0.3$ | | | $3.6 \pm 0.6$ |

[a] Consistent with the companion location relative to the FOV for each instrument, in the above flux ratios the H–band total flux ($F_T$) includes the secondary star, but the K–band total flux does not.

[b] $D_\star = 14 R_\odot$.

[c] Only needed to determine the continuum disk size, not the line sizes.